\def\hii{\ion{H}{2}}

\newcommand{\beq}{\begin{equation}}
\newcommand{\eeq}{\end{equation}}
\documentclass[11pt,preprint]{aastex}
\shortauthors{GARNETT, GALARZA, AND CHU}
\shorttitle{LMC HE II NEBULA N44C}
\begin{document}
\title{THE HE II EMITTING NEBULA N44C IN THE LMC: OPTICAL/UV 
SPECTROSCOPY OF THE NEBULA AND ITS IONIZING STAR\altaffilmark{1}}
\author{Donald R. Garnett\altaffilmark{2}}
\affil{Steward Observatory, University of Arizona, 933 N. Cherry Ave., 
Tucson, AZ  85721 \\ E-mail: dgarnett@as.arizona.edu}
\author{Vanessa C. Galarza}
\affil{Astronomy Department, New Mexico State University, Las Cruces,
NM 88003-8001; \\ e-mail vgalarza@nmsu.edu} 
\and
\author{You-Hua Chu\altaffilmark{2}}
\affil{Astronomy Department, University of Illinois, 1002 W. Green Street,
Urbana, IL 61801;\\ e-mail: chu@astro.uiuc.edu}
\altaffiltext{1}{Based in part on observations with the NASA/ESA Hubble Space
Telescope obtained at the Space Telescope Science Institute, which is operated
by the Association of Universities for Research in Astronomy, Inc., under NASA
contract NAS5-26555.}
\altaffiltext{2}{Visiting Astronomer, Cerro Tololo Interamerican Observatory,
National Optical Astronomy Observatories, which is operated by the Association
of Universities for Research in Astronomy, Inc., under cooperative agreement 
with the National Science Foundation.}

\begin{abstract}
We present {\it HST} spectroscopy and imaging, along with new ground-based 
spectroscopy and {\it ROSAT} HRI imaging, of the He~II emitting nebula N44C 
and its ionizing star. A GHRS spectrogram of the ionizing star yields a spectral
type of about O7 for the star. The lack of P Cygni profiles for Si IV and C~IV 
indicates that the star is not a supergiant. The nebular abundances in the 
ionized gas are consistent with average abundances for LMC H~II regions, with 
the possible exception that nitrogen may be enhanced. Enrichment by a former
evolved companion star is not evident. A long-slit echelle spectrogram in 
H$\alpha$ + [N~II] shows no evidence for high-velocity gas in N44C. This 
rules out high-velocity shocks as the source of the nebular He~II emission.
A 108 ks ROSAT HRI image of N44C shows no X-ray point source to a 3$\sigma$
upper limit $L_X$ $<$ 10$^{34}$ erg s$^{-1}$ in the 0.1-2.0 keV band. Based
on new measurements of the electron density in the He~II emitting region, we 
derive recombination timescales of $\approx$ 20 yrs for He$^{+2}$ and $\approx$ 
4 yrs for Ne$^{+4}$. If N44C is a fossil X-ray ionized nebula, this places 
severe constraints on when the putative X-ray source could have turned off. 
The presence of strong [Ne~IV] emission in the nebula is puzzling if the
ionizing source has turned off. It is possible the system is related to
the Be X-ray binaries, although the O star in N44C does not show Be 
characteristics at the present time. Monitoring of X-rays and He~II emission
from the nebula, as well as a radial velocity study of the ionizing star, 
are needed to fully understand the emission line spectrum of N44C. 
\end{abstract}

\section{Introduction}

Despite more than a century of observing and classifying stars, nature 
continues to surprise us with exotic and fascinating new species in 
the stellar zoo. Such has been the case with the recent discovery of 
several otherwise normal H~II regions that exhibit strong nebular He~II 
recombination line emission 
\citep{sth86, pa86, gkcs91}. 
Such emission is unknown in normal H~II regions: He$^{+2}$ requires 
ionizing photons of at least 54 eV to exist in photoionized nebulae, 
and so is normally seen only in nebulae with very hot stars such as 
planetary nebulae. Nevertheless, six H~II regions with nebular He~II 
emission have been identified in Local Group galaxies (see 
\citet{gkcs91, dlmr90}). 
Four of these nebulae appear to be ionized by high-excitation WO or 
WN stars, while one is ionized by the massive X-ray binary LMC X-1 
\citep{pa86}. 
The sixth, the nebula N44C in the LMC, has an uncertain ionization source, 
and is the subject of this paper. 

Of all these objects, LMC-N44C is perhaps the most
intriguing. N44C is part of the large H~II complex N44, which includes 
the OB associations LH 47 and LH 48. N44C is a very high-excitation 
nebula, with [O~III] 5007 \AA/H$\beta$ ratios as high as 8 in parts 
of the nebula, and strong He~II emission as well 
\citep{sth86} -- 
characteristics common to planetary nebulae. 
Nevertheless, N44C is too large ($\approx$ 60 arcsec or 14.5 pc 
diameter) and too luminous [L(H$\beta$) $\approx$ 4.5$\times$10$^{37}$ 
ergs s$^{-1}$, L($\lambda$4686) $\approx$ 6$\times$10$^{35}$ ergs s$^{-1}$ 
\citep{gkcs91}]
to be a classical PN. From the observed nebular 
L($\lambda$4686)/L(H$\beta$) luminosity ratio, 
\citet{sth86} and 
\citet{gkcs91} 
both deduced that an ionizing star 
would need to have $T_{eff}$ $>$ 70,000 K, based on non-LTE model stellar 
atmospheres, to account for the observed He~II emission. 

Photometric studies of stars in the field of N44C 
\citep{sth86,om95},
show only one likely candidate for the ionizing star --
identified as Star 2 in 
\citet{sth86},
with V = 14.2, B--V = --0.22 \citep{om95}. 
Yet this star appears to be a normal O star: estimates of the 
spectral type of this star range between O4 and O7 
\citep{sth86,pm89,om95}. 
Spectral classification of the star is somewhat uncertain because of 
contamination from strong emission lines of H, He~I, and He~II from 
the surrounding high surface brightness nebula.

Determining the spectral type of the star is critical to understanding the 
nature of the ionization source. A normal mid to late-type O main sequence 
star is very unlikely to produce sufficient photons above 54 eV to account 
for the nebular He~II emission. Because of this, \citet{pm89} 
hypothesized that N44C is a fossil X-ray-ionized nebula whose X-ray source
had switched off in the recent past. On the other hand, non-LTE stellar 
atmosphere models by 
\citet{ggkm92} 
suggest that hot ($T_{eff}$ $>$ 
50,000 K) stars near the Eddington limit could account for observed 
nebular He~II emission in some cases; such stars would appear as extreme O 
supergiants. Although the published results appear to rule out this 
hypothesis for the ionizing star of N44C, the uncertainty in determining the 
spectral type from the optical spectrum suggests that another approach is 
in order. 

In this paper we present a comprehensive imaging and spectroscopic study 
of N44C and its ionizing star. We have obtained narrow-band imaging and
UV/optical spectroscopy of the nebula using both ground-based telescopes
and the {\it Hubble Space Telescope} to study the ionization balance and 
element abundances in the nebula, in order to better understand the 
ionization mechanism and to look for abundance anomalies that would signal 
the presence of a highly evolved star. We have also obtained an ultraviolet 
spectrogram of Star 2 with the Goddard High Resolution Spectrograph (GHRS) on 
{\it HST} to determine the stellar spectral type free of uncertainties due 
to nebular contamination. Our goals are to understand better the physical 
properties of N44C and the ionizing star, and to answer the question of
whether Star 2 is the actual source of the ionizing photons powering the 
nebular He~II emission. 

The importance of understanding the origin of this unusual nebular He~II
emission goes beyond the peculiarity of the stars themselves. Numerous 
metal-poor starburst galaxies exhibit narrow He~II emission (e.g., 
\citet{ctm86,ks83}), 
often interpreted as
nebular emission. This may indicate that unusually hot ionizing stars are 
relatively common in starbursts. Many do not show WR features however, and 
so hot O stars may be the cause. The existence of very hot extreme O stars 
and WR stars in distant giant H~II regions will affect the nebular ionization 
balance and complicate the use of nebular diagnostics to understand 
the massive star population in distant star-forming regions. These hot 
stars may be important in young 
galaxies as well, and so may have an important effect on the spectral 
appearance of star-forming galaxies at high redshift. Some authors 
(e.g., \citet{mh91}) 
have identified the He$^{+}$-ionizing stars as the 
putative ``warmers'', postulated hot Wolf-Rayet stars which are capable of 
producing the hard ionizing continuum needed to explain the spectra of some 
active galactic nuclei; synthetic spectra of starbursts including such hot 
stars are also becoming common 
(e.g., \citet{slg92}). 
However, the local examples of these stars need to be studied in greater 
detail before their applicability to AGNs and starbursts can be properly 
assessed.

\section{Observations }

\subsection{Ground-based Optical Imaging}

Narrow-band optical images of N44C were obtained as part of a general search 
for high-excitation H~II regions in the Magellanic Clouds. These images 
were taken in December 1991 using the 0.9-m telescope at Cerro Tololo 
Interamerican Observatory. A Tektronix 1024$\times$1024 CCD was used as the 
detector; the field covered in each image was 405$\arcsec\times$405$\arcsec$, 
at a pixel scale of 0\farcs4 per pixel. The seeing at the time of the 
observations was about 2$\arcsec$.  We used narrow-band filters from the 
Rutgers Fabry-Perot order-sorting set to image the region in H$\beta$, 
[O~III] 5007 \AA, He II 4686 \AA, and the continuum at 5081 \AA. Central 
wavelengths of the four filters were 4861 \AA, 4993 \AA, 4685 \AA, and 
5081 \AA; all filters have bandwidths of 44 \AA\ FWHM. Generally, the 
exposures were short because of the survey nature of the observations: 
600 s for H$\beta$, 300 s each for the [O~III] and 5081 \AA\ filters, and 
900s in the 4686 \AA\ filter.

For each image, we subtracted a DC offset determined from the overscan
region of each frame, as well as a bias which was constructed from the 
median of 35 zero exposure images. The images were flat-fielded using 
exposures of the illuminated dome taken through the appropriate filter. 
Seven to ten dome exposures were obtained through each filter and combined 
by averaging with a sigma-clipping algorithm to reject cosmic rays. The 
combined images were normalized to provide a master flat field for each 
filter, which was then divided into the object frames. We converted 
count rates in the reduced emission line images to fluxes using our 
long-slit 4-m spectra (see below). We will use these images primarily to 
measure integrated H$\beta$ and He~II 4686 \AA\ luminosities for N44C.

\subsection{Ground-based Optical Spectroscopy}

Long-slit optical spectra of N44C were also obtained at CTIO in December 
1991. We used the RC spectrograph on the 4-m reflector to obtain both blue 
and red spectra of the nebula. Grating KPGL1 was used in first order to 
observe the 3350-5500 \AA\ wavelength region at 3.8 \AA\ resolution (FWHM), 
while grating 181 was used in first order with a GG385 order-sorting 
filter to cover the 4200-8500 \AA\ region at approximately 7 \AA\ 
resolution; a Reticon 1200$\times$400 CCD was the detector. Five 120 s
exposures were made with the blue setting and seven 60 s exposures with the 
red setting; the exposure times were kept short to avoid overexposure in the 
very strong [O~III] lines. The 2$\arcsec$ by 330$\arcsec$ slit was oriented 
N-S and approximately centered on Star 2; both Star 2 and Star 1 located 
about 8$\arcsec$ to the south were included in the slit. 

Our reduction of the data followed common practice for two-dimensional
spectra. The DC offset and a bias image were subtracted from the spectral 
images. A normalized flat-field, constructed from images of the illuminated 
dome and twilight sky, was divided into each frame to correct for both 
pixel-to-pixel sensitivity variations and non-uniform illumination of the 
slit. We solved for the wavelength as a function of pixel position using a 
He-Ne-Ar arc lamp spectrum, fitting a low-order polynomial function to the
line positions. The rms residuals of the actual to fitted 
wavelengths were of order 0.1 pixel, or 0.4-0.7 \AA. The spectra were 
corrected for atmospheric extinction using mean CTIO extinction coefficients. 
Finally, the photometric flux calibration was determined from observations of 
spectrophotometric standard stars from the list of 
\citet{sb83}, 
with the red extended data of 
\citet{hamuy92}. 
The standard stars were observed with 
the slit width opened to 10$\arcsec$ to reduce slit losses. At least three 
standards were observed on each night. The calibrated images were combined 
with an averaged sigma-clipping algorithm.

We extracted a one-dimensional spectrum by summing the two-dimensional data
over the entire region of emission from N44C, including the ionizing star, 
and excluding the nebular emission outside the nebular boundary. 
Our spectra were taken during near full moon, so the sky background was 
relatively high. Background subtraction for the N44C frames was complicated 
by the extent of the nebulosity along the slit and by numerous stars within 
the slit. Nevertheless, we located sufficient regions clear of nebulosity 
and stars to determine and subtract accurately the sky background. The 
background was determined by a linear fit over those regions. Residual 
[O~I] 5577 \AA\ night-sky emission in the red spectrum amounted to 
only about 3\% of the original line flux, while residual OH sky emission 
was not detected. The Hg I 4358 \AA\ airglow line was almost negligibly 
weak in our blue spectrum; the [O III] 4363 \AA\ line, shifted to 4368 \AA\ 
by the LMC radial velocity, was thus not affected by airglow contamination. 
Figure 1 displays our fully calibrated and extracted one-dimensional blue 
and red spectra.\notetoeditor{WE PREFER THAT FIGURE 1 SPAN TWO COLUMNS}

The Reticon spectra suffered somewhat from focus variations due to the 
wrinkled surface of the chip. Care was taken so that important closely-separated 
sets of lines, such as H$\gamma$ and [O~III] 4363 \AA\ and the 
H$\alpha$-[N~II] group, were in well-focused regions of the detector. The 
effects of the focus variations are evident in the distorted profiles of 
the [O~III] 4959, 5007 \AA\ and H$\beta$ lines in the blue 
spectrum (Figure 1). 

\subsection{HST Spectroscopy}

N44C-Star 2 was observed with the Goddard High Resolution Spectrograph 
(GHRS) on the {\it Hubble Space Telescope} during Cycle 6. We employed the 
low-resolution grating G140L to observe the 1200-1765 \AA\ region in two 
settings with a resolving power R = $\lambda/\Delta\lambda$ $\approx$ 
2000. A journal of the observations is given in Table 1. A loss of guide 
star lock resulted in a premature end to the shorter-wavelength GHRS 
observation, with central wavelength 1342 \AA, from the planned 760 s 
to 235 s. Despite this, the spectrum had sufficient signal/noise 
(S/N $\approx$ 10 per pixel) for 
classification of the star. The redward exposure, with central wavelength 
1617 \AA\, was carried out with the planned exposure time. The spectra were 
reduced by the STScI pipeline. Comparison of the blue and red spectra
showed excellent photometric consistency in the 18 \AA\ 
region of overlap. We conclude from this that the loss of lock did not 
affect the photometry of the first spectrum. We show the merged GHRS 
spectrogram of Star 2 in Figure 2, along with UV spectra of comparison
O III-V stars in the Magellanic Clouds from 
\citet{wal95}.
\notetoeditor{PLEASE MAKE FIGURE 2 A FULL PAGE PLOT IN LANDSCAPE MODE}

Immediately following the GHRS observations, we obtained spectroscopy of 
the nebula with the Faint Object Spectrograph (FOS). The 0\farcs86
circular aperture was positioned 3$\arcsec$ north of Star 2, to sample
the region of highest surface brightness. The FOS observations were
planned to cover the 1200-6800 \AA\ spectral region with five of the 
high-dispersion (R = 1200) gratings. The two 
original FOS/BL observations failed because the FOS aperture door 
failed to open during the exposures. We therefore repeated four of 
the FOS exposures in February 1997 (excluding G270H). Comparison of the 
G190H exposures taken at the two epochs, and the G570H exposures also, 
showed good agreement. Therefore, we averaged the multiple G190H and G570H 
exposures to improve the signal/noise. These data were also processed with 
the STScI pipeline, and Table 1 gives the journal for these observations. 
Portions of the FOS spectrogram are shown in Figures 3 and 4.
\notetoeditor{PLEASE MAKE FIGURES 3 AND 4 SPAN TWO COLUMNS}

\subsection{WFPC-2 Imaging}

In addition to the {\it HST} spectroscopy, we obtained WFPC-2 narrow-band
images of N44C in Cycle 6 to look for evidence of filamentary structure
that might indicate a shock and to look for evidence of a companion to
Star 2 on scales of a few thousand AU (0.1 arcsec $\approx$ 5000 AU at
the distance of the LMC). We collected images in F502N ([O III]), F547M
(emission-free continuum), F656N (H$\alpha$), and F675W (red continuum).
The telescope was positioned so that Star 2 was centered in chip WF2 of
the camera. These images were processed through the HST pipeline and 
mosaicked to produce the final images. A color composite of the F656N, 
F502N, and F547M images is shown in Figure 5.
\notetoeditor{PLEASE MAKE FIGURE 5 A FULL-PAGE FIGURE}

\section{The UV Spectrum of the Ionizing Star}

N44C Star 2 was identified as the primary source of ionizing photons for 
the nebula early on by 
\citet{sth86}, 
based on the stellar colors,
O-type spectral features in the optical spectrum, and the fact that the
He~II 4686 \AA\ nebular emission is centered on this star. Another
bright star some 8 arcsec south of Star 2, Star 1 in 
\citet{sth86}, 
with magnitude V = 13.8 and color B-V = 1.06 
\citep{sth86}, 
shows strong features of Ca~I 4227 \AA, the G-band of CH, and the Mg~I b lines 
in our CTIO spectrogram. These features clearly indicate that Star 1 is a 
late-type star, most likely a foreground G or K star. 

Our purpose in obtaining the GHRS spectrogram was to determine an accurate
spectral type for Star 2 and to look for spectral peculiarities. Spectral
typing of O stars embedded in bright nebulosity is sensitive to the 
subtraction of the strong, spatially variable H and He recombination lines
from the nebula, and this is true for N44C. In contrast, the UV spectrum
of O stars includes a number of spectral features which are sensitive
to spectral type and luminosity 
\citep{wnp85, wal95}, 
while nebular emission is largely
absent in the UV. For example, the N~V 1240 \AA, N~IV 1718 \AA, 
and O~V 1371 \AA\ features provide some discrimination between early- 
and late-type O stars, while the Si~IV 1400 \AA\ feature strongly 
discriminates between supergiants and dwarfs, developing a distinctive 
P Cyg wind profile in O supergiants of type O4 and later. 

Our GHRS spectrogram displayed in Figure 2 shows features typically found in 
O stars, the most prominent of which include N V 1238,42, C IV 1548-50,
Si IV 1398, 1420, N IV 1718, He II 1640, and O IV 1339,44. The N V feature
exhibits an unsaturated  P Cygni profile characteristic of a strong stellar 
wind, but except for a possible weak C IV feature, no other stellar feature 
has a P Cyg profile at the limits of the signal/noise of the spectrogram. 

The spectrum also shows a number of strong, narrow interstellar absorption
features. The most prominent is the saturated Ly $\alpha$ line (with the
geocoronal Ly$\alpha$ emission in the core; note also that the Ly$\alpha$
absorption goes to zero, indicating that scattered light is absent). The 
centroid of the absorption
line is offset from the geocoronal emission by about 250 km s$^{-1}$, 
evidence that the bulk of the absorbing neutral hydrogen lies within the
LMC. Other prominent interstellar features include the S II/C I blend at
1260 \AA, O I + Si II near 1304 \AA, C II 1335 \AA, Si II 1526 \AA, and
Al II 1670 \AA. Most of these lines show double profiles separated by
approximately 230 km s$^{-1}$. The low velocity component likely arises
from foreground gas in the Galaxy, while the higher velocity component
arises from gas in the LMC. 

\citet{ggkm92} 
proposed that extreme O supergiants with $T_{eff}$ $>$ 
50,000 K close to the Eddington limit could produce sufficient numbers 
of He$^+$-ionizing photons to account for nebular He~II emission in some 
cases. On the other hand, nebular He~II recombination emission has not 
been seen associated with Population I O3 stars in either the Galaxy or 
the Magellanic Clouds.
Our GHRS spectrogram shows clearly that Star 2 is not an early O
supergiant. Such stars in the Galaxy and the Magellanic Clouds display 
saturated P Cygni profiles for C~IV 1550 \AA\ and N~V 1550 \AA, as well
as P Cyg profiles for He~II 1640 \AA\ and N~IV 1718 \AA. Supergiants of
type O4 and later also show P Cyg profiles for Si~IV 1400 \AA. Except for
N~V, none of these lines shows a prominent P Cyg profile in the spectrum 
of Star 2. In addition, neither our low-resolution optical spectrum nor
a high-dispersion echelle spectrum (kindly provided by K. Venn and D. 
Lennon) for Star 2 show the N~IV 4058 \AA\ feature nor the N~V 4604, 4620 
\AA\ lines observed in O3-O4 supergiants. 

At the same time, the UV spectrum shows some evidence for peculiarities
that confuse the spectral classification. Comparing with the O5 V star
Sk--70 69 and the O9 III star AV 238 (Figure 2), we see that while 
N~V 1240 \AA\ has a strength consistent with the earlier spectral type,
the C~IV feature has a strength more similar to the later spectral type.
The actual strength of the C~IV profile is difficult to determine because
of the likely presence of superposed C~IV interstellar absorption from 
both the Galaxy and the LMC. However, the stellar C~IV feature is noticeably
weaker than observed in other O7 V stars 
\citep{wnp85, wal95}). 
The strength of N~V and the weakness of C~IV suggest the possibility
that Star 2 is an ON-type star 
\citep{wp85}. 
Enhanced N and deficient C in the star would indicate that it is moderately 
evolved. A more detailed analysis is needed to determine if there are actual 
abundance anomalies in the star. Nevertheless, it is clear that Star 2 is an 
unlikely source for the photons capable of producing the nebular He~II emission.
On the basis of the UV spectrum, we estimate the spectral type of Star 2 
to be approximately O7 III-V, consistent with that derived by 
\citet{om95} 
from the optical spectrum. 

\section{Features from Narrow-Band WFPC2 Imaging}

The composite WFPC2 image of the N44C region (Figure 5) shows three
main features: (1) the high surface brightness, high-excitation 
([O~III] 5007/H$\beta$ $\sim$ 6) main body of N44C, which is mostly
contained in chip WF2 (upper left); (2) diffuse, high-excitation
filamentary emission to the SW of N44C; and (3) low-excitation
diffuse and dusty gas, seen mostly as H$\alpha$ emission to the N 
and W of the bright \hii\ region.

The bright main body of N44C has the appearance of a normal \hii\
region, apart from its high excitation and He~II emission. The E,
NE, and NW boundaries of N44C are relatively sharp and have low 
[O~III]/H$\alpha$, indicating the presence of an ionization front. 
We infer that the nebula is bounded by a molecular cloud on these
three sides. The low-excitation H$\alpha$ emission to the N of the
nebula, separated by a dust lane from N44C, is part of the superbubble
surrounding the association LH 47. 

The WFPC2 images shows Star 2 to be single to the limits of the
WFPC2 resolution, about 5000 AU at a distance of 50 kpc. This is
consistent with the normal appearance of the spectrum of Star 2.
Any companion must be much fainter than the O star, unless it is
of the same spectral type. 

A large region of diffuse filamentary gas is seen to the SW of N44C.
These filaments appear to outline a roughly spherical cavity, and
have [O~III]/H$\alpha$ comparable to what is seen in N44C. The 
streamers running south from N44C may be part of this structure. 
It is clear that these streamers are foreground to N44C since they
can be seen in dust absorption against the bright western part of
the nebula. We therefore infer that the diffuse filamentary structure 
is physically unrelated 
to N44C. The structure may be a wind-blown bubble associated with 
the Wolf-Rayet star Br 25, which can be seen at the bottom of the PC 
frame in Figure 5. Such bubbles are known to have high-excitation
shells (e.g., NGC 2359; \citet{duf89}). 

Perhaps more intriguing is the series of east-west oriented arcs
located approximately 40$\arcsec$-50$\arcsec$ south of Star 2. The arcs
are concave towards N44C, giving the impression of an expansion in
that direction, possibly associated with a stellar outburst. Our
echelle spectrogram slit (discussed below) did not cross the arcs. 
However, the slit used by \citet{gm84} in 
their kinematic study did cross a part of these arcs. As discussed
below, they saw little evidence for high-velocity gas in N44C, apart
from the feature we now know to be He~II 6560 \AA. In the absence
of stronger kinematic evidence, we speculate that the brightest arc 
represents an edge-on ionization front, isolated from the main 
nebula by a superposed dust lane and/or a shadow region. 

The PC frame shows a region of dusty material. This region is most
likely associated with the star-forming complex N44B, which is located
just outside the image to the NW (upper right). This region shows 
extensive dust structures and also extensive nebular emission for 
which the ionizing sources are not obvious. Infrared imaging of this 
region could reveal evidence of significant star formation activity.

\section{The Nebular Spectrum and Element Abundances}

We measured the emission-line features in each spectrogram using the SPLOT 
routine in IRAF\footnotemark. 
\footnotetext{IRAF is distributed by the National Optical Astronomy 
Observatories, which are operated by the Association of Universities for
Research in Astronomy, Inc., under cooperative agreement with the National
Science Foundation.}
Line strengths relative to H$\beta$ were measured by integrating the
fluxes under the emission-line profiles. 
The hydrogen Balmer emission line strengths in the CTIO spectrogram were 
corrected for stellar absorption by the ionizing star using an approximate 
spectral type of O7, based on the initial inspection of the HST GHRS stellar 
spectrogram. This corresponds to an absorption equivalent width of 2.8 \AA\ 
for the H$\beta$ line. Certain features, such as [Ar~IV] 4711 \AA\ + 
He~I 4713 \AA, are strongly blended. In such cases, we determined the
relative fluxes for the individual lines by using the appropriate 
theoretical ratios for hydrogen and helium from 
\citet{hs87}. 
The measured fluxes for strong lines 
in the overlap region of the red and blue CTIO spectra agreed to within
5\%, so we averaged the measurements in the overlap region.
Upper limits for some undetected lines were determined from the rms noise 
in the local continuum. 

\subsection{Reddening Corrections}

Assuming an intrinsic H$\alpha$/H$\beta$ ratio equal to 2.86 for $T_e$ = 
10,000 K and $n_e$ = 100 cm$^{-3}$, close to the derived physical conditions 
for N44C (see below), we estimate that the interstellar reddening toward 
N44C corresponds to $A_V$ = 0.25$\pm$0.05 magnitude, using a standard 
Galactic extinction curve 
\citep{ccm89}
with $R_V$ = 3.1. This is roughly consistent with the foreground extinction 
toward the LMC of $A_V$ = 0.28 magnitude, based on the extinction maps of 
\citet{bh84}. The fact that we derive slightly 
smaller reddening may be attributed to uncertainties in the reddening 
measurement, and to patchiness in the foreground extinction which is not
resolved by the large beam of the Burstein \& Heiles maps. 

Correcting the FOS emission line measurements for interstellar reddening 
is less straightforward, as the UV reddening function can vary dramatically 
with environment 
\citep{ccm89}. 
Even within the LMC there is no `standard' UV reddening curve, as 
demonstrated by 
\citet{f86}, 
who showed that
the UV reddening curve differs for stars inside and outside the 30 Doradus
region. Our FOS spectrogram has a number of He~II and He~I recombination lines 
(in particular the bright He~II 1640 \AA\ line) in 
the UV and visible spectral regions, which we could use to constrain the UV 
reddening curve. We compared the theoretical ratios for He II lines with
respect to He~II 4686 \AA\ and He I lines with respect to 4471 \AA,
along with the hydrogen Balmer decrement, with reddening-corrected ratios 
based on the LMC reddening curves derived by 
\citet{f86}
and the standard Galactic reddening curve with $R_V$ = 3.1 from 
\citet{ccm89}. We
found empirically that the average interstellar reddening curve for regions 
within 0.5 degrees of 30 Dor derived by 
\citet{f86}, 
with $A_V$ = 0.26$\pm$0.05 mag, provided the best match to the relative 
strengths of the He~II, He~I, and H I lines. 

The reddening-corrected line fluxes for the CTIO spectrogram are listed in 
Table 2, while those for the FOS spectrogram can be found in Table 3. The
tabulated fluxes are normalized to the H$\beta$ flux in both tables. The
tabulated uncertainties in the line fluxes were evaluated from the rms noise 
in the continuum near each emission line, combined with the uncertainty in 
the relative photometric calibration across the spectrum (estimated to be 
approximately 3\% based on standard stars) and the uncertainty in the 
reddening correction.

\subsection{Physical Conditions}

It is necessary to determine the electron temperature $T_e$ and electron 
density $n_e$ in the ionized gas in order to derive ionic abundances from 
the emission-line spectrum. A number of diagnostic emission line ratios 
are measureable in our spectra (Table 3). We used an updated version of 
the Lick Observatory five-level atom program FIVELEV 
\citep{ddh87} 
to derive the physical conditions from the 
diagnostic line ratios. The various diagnostic emission-line ratios and 
physical conditions derived from the line ratios are listed in Table 5. 

The [S~II] line ratio integrated over the CTIO spectrogram corresponds to $n_e$ 
in the low-density limit. However, the [Ar~IV] line ratio in the CTIO spectrogram 
and the [S~II] and [Ar~IV] ratios in the FOS spectrogram yield a higher density 
of approximately 2000 cm$^{-3}$ in the high surface brightness region within 
3$\arcsec$ of the ionizing star. Thus, the bulk of the nebula, as sampled
by the larger CTIO slit, appears to consist of moderately low-density material, 
with the central He~II-emitting region sampled by the FOS aperture being the 
illuminated face of a molecular cloud in close vicinity to 
the ionizing star, similar to what is observed in the Orion Nebula 
\citep{bf91}. 
The higher density in this central region is still 
too low to have a significant effect on the abundances derived for the nebula.

The various results for $T_e$ from [O III] line ratios in Table 3 agree to 
within measurement errors. A weighted mean of the three measurements yields 
$T_e$ = 11,200 $\pm$ 200 K. We adopt this value for the ion abundance calculations 
to follow. A plot of the 5007 \AA/4363 \AA\ ratio along the CTIO 
slit shows little variation other than noise, suggesting that the nebula is 
largely isothermal.
The values of $T_e$ derived from the [S~II] 4069 \AA/6725 \AA\ and 
[O~II] 2470 \AA/3727 \AA\ ratios are somewhat smaller, approximately 
9000 K; the uncertainties are large, however, and the results are sensitive
to the assumed electron density. The better measurement of [O~II] 
3727 \AA/7325 \AA\ yields a temperature much closer to the [O~III]
temperature. Although photoionization models 
(e.g., \citet{g92}) 
predict that 
the electron temperature in the low-ionization zone (including O$^+$, N$^+$, 
and S$^+$) should be about 400 K cooler, we will treat the nebula as isothermal 
here. Only a small error in the abundances derived here will be introduced by 
such an assumption, since the nebula is highly ionized. As is shown in the next 
section, the ionic abundances determined from the CTIO spectra and the FOS 
spectra are in good agreement with each other, indicating that the isothermal 
assumption does not lead to very large errors. We adopt $T_e$ = 11,200 K and 
$n_e$ = 150 cm$^{-3}$ for the CTIO observations, and $T_e$ = 11,200 K, $n_e$ =
1500 cm$^{-3}$ for the FOS observations. 

\subsection{Ionic Abundances}

We calculated ionic abundances relative to H$^+$ from each spectrum separately
using the five-level atom code, adopting the $T_e$ and $n_e$ values listed in 
the previous section. In general, the brightest or least density-sensitive 
measurable emission lines of each ion were used to determine the ionic abundances. 
The abundances for all ions observed in both spectra are listed in Table 3.

The He$^{+}$ and He$^{+2}$ abundances were determined using the theoretical
emissivities from 
\citet{bss99}, 
including the contributions for collisional excitation of He~I. 
We used $n_e$ $\approx$ 1500 cm$^{-3}$ and $T_e$ = 11,200 K to derive
He$^+$ from the FOS spectrogram; at this electron density, the  
collisional contribution is at most 15\%, for the 5876 \AA\ line, and 
less for the other He~I lines. For the CTIO observation, we used $n_e$ = 
150 cm$^{-3}$ to compute the He$^+$ abundance; at the higher density 
used for the FOS measurements, we obtained larger differences between the 
He$^+$ abundances derived from the 5876 \AA, 4471 \AA, and 6678 \AA\ 
lines than when we use the lower density. The He$^+$ abundance from
the CTIO data is a weighted mean of the values from the 4026, 4471, 5876,
and 6678 \AA\ lines; for the FOS data, we used the weighted mean values
from the 2945, 3188, 4471, 5876, and 6678 \AA\ lines. For He$^{+2}$, we
used the 4686 \AA\ line only. 

The ionic abundances determined from the two spectral datasets are in
relatively good agreement, apart from differences that can be attributed
to variations in ionization. The small He~II-bright region observed 
within the FOS aperture exhibits higher ionization than the larger region 
encompassed by the CTIO slit, as reflected by the larger derived abundances 
for He$^{+2}$ and Ar$^{+3}$ and the lower abundances for O$^+$, N$^+$, and
S$^+$ from the FOS spectrogram. The O$^{+2}$ and Ne$^{+2}$ abundances are
roughly the same between the two spectra, but these two species would be
expected to be partially replaced by O$^{+3}$ and Ne$^{+3}$ in the He$^{+2}$
zone, so little variation would be expected. 

\subsection{Total Abundances}

\subsubsection{Nebular Parameters and Ionization Models}

The total abundances for each element can be determined, in principle, 
by summing the contributions from each ionization state. For many of the
elements we study here, however, some important ionization states are
not observed, or have poorly determined limits. To estimate corrections
for unseen ionization states, we have computed photoionization models
using the code Cloudy version 90.04 
\citep{cloudy96}. 

First, we establish a few stellar and nebular properties. We estimated 
the total H- and He$^+$-ionizing photon fluxes by determining the H$\beta$ 
and He~II 4686 \AA\ fluxes from our CTIO images. We measured the counts 
in the H$\beta$ and He~II 4686 \AA\ images (with stars subtracted) within 
a 2$\arcsec$ strip centered on Star 2, corresponding to the slit used for the 
CTIO spectra. The H$\beta$ and 4686 \AA\ fluxes measured in our CTIO 
spectrogram were used to determine the conversion from counts to flux in the 
images. We then measured total line fluxes for H$\beta$ and He~II in the 
region encompassing the nebula from the images. The measured fluxes were 
F(H$\beta$) = (3.0$\pm$0.6)$\times$10$^{-10}$ erg cm$^{-2}$ s$^{-1}$ 
and F(4686) = 7.3$\times$10$^{-13}$ 
erg cm$^{-2}$ s$^{-1}$, after correcting for obscuration $A_V$ = 0.25 mag,
corresponding to emission-line luminosities L(H$\beta$) $\approx$ 
9$\times$10$^{37}$ erg s$^{-1}$ and L($\lambda$4686) $\approx$ 
2.2$\times$10$^{35}$ erg s$^{-1}$ at a distance of 50 kpc. For case B 
recombination, the derived H-ionizing photon luminosity is Q(H$^0$) = 
1.9$\times$10$^{50}$ s$^{-1}$ and Q(He$^+$) = 2.3$\times$10$^{47}$ s$^{-1}$. 

The values for L($\lambda$4686)/L(H$\beta$) and Q(He$^+$)/Q(H$^0$) derived
here for N44C are a factor of six smaller than those derived earlier by
\citet{gkcs91} 
from published measurements. There are several 
possible reasons for this discrepancy. First, the line fluxes in 
\citet{gkcs91} 
were derived from the average surface brightnesses obtained 
from the small-aperture spectra of 
\citet{sth86}, 
multiplied 
by the estimated nebular area. These fluxes may have significant uncertainties
due to the large geometric correction and the uncertainty in the average
surface brightness. Second, the He~II flux derived from our CTIO image
may have a large uncertainty arising from the subtraction of the bright
stars in the nebula. We estimate that our value for the $total$ He~II
line flux may be uncertain by $\pm$50\%. Third, it is possible that the
He~II flux may have experienced a significant decline between the 1986
observations and our 1991 observations (see discussion below). 

Following the example of 
\citet{fer98} 
for the Orion Nebula, we can estimate 
the distance from Star 2 to the ionization front for our FOS position from 
the nebular surface brightness and the derived Q(H$^0$). We assume that the
bulk of the H$\beta$ emission within the FOS aperture arises from the surface 
of a molecular cloud
on the far side of the star from us (see section 5). The H$\beta$ surface 
brightness within the FOS aperture is 4.0$\times$10$^{-14}$ erg cm$^{-2}$ 
s$^{-1}$ arcsec$^{-2}$. If the nebula emits isotropically, this corresponds 
to 4$\pi$J $\approx$ 0.02 erg cm$^{-2}$ s$^{-1}$ at the surface, or about 
5.2$\times$10$^9$ H$\beta$ photons cm$^{-2}$ s$^{-1}$. In case B recombination, 
approximately one out of every nine recombinations produces an H$\beta$ photon, 
so the recombination rate at the surface is approximately 4.5$\times$10$^{10}$
cm$^{-2}$ s$^{-1}$. For an ionization-bounded nebula, the ionizing photon
surface flux must be at least equal to the surface recombination rates. 
Thus, for our derived Q(H$^0$), the separation between Star 2 and the 
ionized surface layer is about 1.8$\times$10$^{19}$ cm. In reality, 
photons reflected by the molecular cloud will contribute a small amount
to the H$\beta$ flux, and grains will compete with H atoms for ionizing
photons, which will lead to a somewhat smaller derived separation. For
the purposes of our analysis, we will assume a star-gas separation of
10$^{19}$ cm. 

Ionizing stellar fluxes for the photoionization calculations were taken from 
the NLTE stellar atmosphere models of \citet{hus84}. 
Other more recently computed atmosphere flux models are available, but as
we shall see below the corrections for unseen ionization states are small
for most of the elements we observe, and the uncertainty in the ionizing
radiation field is not critical to our estimates of element abundances
in the case of N44C. We used average LMC element abundances from 
\citet{gar00} and \citet{duf90}, 
with refractory elements depleted by a factor of ten. Orion-type 
grains were included with an abundance of 0.25 times the Milky Way value. Model 
runs were computed for $T_{eff}$ = 50,000, 75,000, and 85,000 K stellar 
atmospheres, and for ionization parameters log U = $-$2.5, $-$3.0, and $-$3.5. 
Estimated total abundances for the elements observed in N44C are listed 
in Table 6. We discuss individual elements in detail below.

\subsubsection{Helium}

Computing the total helium abundance is trivial in principle, being the
simple sum of the He$^+$ and He$^{+2}$ abundances. Simple photoionization
models predict negligible neutral helium in the H$^+$ zone for stars at 
least as hot as O7. 

Table 8 shows that He/H derived from the CTIO and FOS spectra disagree, 
with a 2.5$\sigma$ difference between the two values. This is unlikely
to be due to unaccounted observational errors. One possible explanation
is that the adopted electron density for the FOS observation may be too 
large. The He$^+$ abundance is 
sensitive to the correction for collisional excitation. Lowering the 
electron density would result in a higher He$^+$ abundance. Nevertheless, 
we can conclude that there is no evidence for He enrichment in N44C; 
the CTIO value is in good agreement with the average LMC He/H listed in 
\citet{duf90}.

\subsubsection{Oxygen}

Most of the oxygen in \hii\ regions is in the form O$^+$ and O$^{+2}$.
However, in regions where He$^{+2}$ is present O$^{+3}$ is expected as
well. Our photoionization models indicate that the O$^{+3}$ fraction
is tightly correlated with He$^{+2}$/He$^+$, and depends only weakly
on the ionization parameter. The predicted corrections to 
the oxygen abundance are 2\% and 14\% for the CTIO and FOS spectra,
respectively. These results are consistent with the corrections 
derived by the scheme listed in 
\citet{kb94} 
based on planetary nebula models (4\% and 19\% respectively). 
The values of O/H from the two spectra after these
corrections agree to within 3\%. The mean value, O/H = 2.1$\times$10$^{-4}$,
is consistent with the average O/H for LMC \hii\ regions 
\citep{gar00, duf90}.

\subsubsection{Carbon}

Our 2$\sigma$ upper limit on the C~IV emission leads to a significant
upper limit on the C$^{+3}$ fraction, less than 25\% of the total C/H
obtained from the FOS observations. This is consistent with our 
photoionization models,
which predict that C$^{+3}$/C$^{+2}$ $\approx$ 0.1-0.2 for the observed
He$^{+2}$ fraction. C$^{+4}$ contributes negligibly. We therefore adopt
C/H = (7$\pm$1)$\times$10$^{-5}$ for N44C. This value is within 6\% of
the value derived for 30 Doradus by 
\citet{gar95}. 
We thus find no evidence that carbon is enriched in N44C. 

\subsubsection{Nitrogen}

The nitrogen abundance has the largest potential uncertainties, because
only one ionization state, N$^+$, is observed directly. The large 
O$^{+2}$/O$^+$ ratio in N44C tells us that N$^+$ is only a minor fraction
of the nitrogen abundance, and so a large ionization correction will
be incurred.  

The analysis by 
\citet{gar90}
suggested that, for metal-poor \hii\ regions
ionized by stars hotter than 40,000 K, the N$^+$/O$^+$ ratio was a good
approximation for N/O, to within 20\% accuracy. The ionization models
computed here confirm those results. We find from the models that 
(1) the ratio N$^+$/O$^+$ varies between 1.0 and 1.15 times N/O, with 
no systematic variation with O$^{+2}$/O$^+$; and (2) the ratio
(N$^{+3}$ + N$^{+2}$)/N$^+$ = O$^{+2}$/O$^+$ to within a few percent
over the entire range of ionization parameter and $T_{eff}$ considered
here. Our 2$\sigma$ upper limit on N$^{+2}$/H$^+$ from the FOS spectrum
is consistent with these results, but is too large to provide a useful
constraint. Using these theoretical relations yields N/O = 0.05 for the
CTIO specturm and N/O = 0.08 for the FOS spectrum. For the rest of our
discussion we adopt the mean of the two results, log N/O = $-$1.2$\pm$0.1,
noting the potentially large uncertainty in the ionization correction.
This value is 0.3 dex larger than the mean value for LMC \hii\ regions
\citep{gar00}. 
Thus there is some evidence for an nitrogen enrichment in N44C. 

\subsubsection{Sulfur, Neon, and Argon}

The neon abundance can be derived straightforwardly from the sum of 
Ne$^{+2}$ and Ne$^{+3}$ from the FOS spectrum, while the Ar abundance
is the sum of Ar$^{+2}$ and Ar$^{+3}$ in the CTIO spectrum. The low
upper limits on Ne$^{+4}$ and Ar$^{+4}$ indicate that these species
contribute negligibly, and the ionization models concur that these
two species are less than 2\% of the total Ne and Ar abundances. The
contributions from Ne$^+$ and Ar$^+$ are predicted to be less than
10\% of the total by the ionization models. 

The total Ar/H abundance for N44C is within the range observed for
ionized nebulae in the LMC and in other galaxies 
\citep{gar00}.
The value log Ar/O = --2.3 in N44C is also close to the mean value
for the LMC. On the other hand, Ne/O is about 0.2 dex higher than 
the mean for LMC \hii\ regions. We are puzzled as to the source of 
the high Ne. We obtain a high Ne abundance even if we consider only
Ne$^{+2}$. One possible solution is if $T_e$ in the regions where 
Ne$^{+2}$ and Ne$^{+3}$ are produced is approximately 13,000 K, 
instead of the 11,400 K used here. This might be expected for such
high-ionization species. On the other hand, $T_e$ for Ne$^{+2}$ is
expected to be similar to that for O$^{+2}$ 
\citep{g92}, 
and one
might also expect to measure a higher $T_e$ in the FOS data than
in the CTIO data, which is not observed. A higher $T_e$ in the 
high-ionization zone would also lead to a significant reduction in
the Ar abundance, leading to a discrepancy with the mean LMC abundance
for that element. A measurement of the high-excitation [Ne~IV] lines
would provide a constraint on $T_e$ for the higher ionization states.

We detect S$^+$ and S$^{+2}$ in both our spectra. For a hot ionizing 
star, significant amounts of S$^{+3}$ and S$^{+4}$ may also be
present. Our photoionization models predict that these two species 
should have abundances of 0.2$\pm$0.1 times the S$^{+2}$ abundance
in the CTIO data and 0.35$\pm$0.1 times the S$^{+2}$ abundance
in the FOS data. The resulting total S abundance for the two
sets of observations are in excellent agreement. Our derived values 
log S/H = --5.4$\pm$0.1 and log S/O = --1.7$\pm$0.1 are consistent 
with the mean value for LMC \hii\ regions. 

\section{Echelle Spectrum}

A long-slit echelle observation of the H$\alpha$ + [N II] lines was made with 
the 4m telescope at Cerro Tololo Inter-American Observatory on 7 January 1988.  
The observing configuration of the spectrograph has been described by 
\citet{ck94}. 
Thorium-Argon lamp scans were used for wavelength calibration and 
distortion correction; however, the geocoronal H$\alpha$ component at zero 
observed velocity was used to fine-tune the velocity calibration. N44C was 
observed with a slit width of 250 $\mu$m (1\farcs64). The resulting 
instrumental FWHM of the H$\alpha$ line was 18$\pm$1 km s$^{-1}$. The slit 
was oriented E-W and passed through Star 2, extending from 175 arcsec east 
of Star 2 to 90 arcsec west of the star. Figure 6 displays a portion of the 
two-dimensional echelle spectrogram. The three 
most prominent lines are [N~II] 6548, H$\alpha$, and [N~II] 6583 from left to
right. The spectrum of Star 2 is the continuous horizontal band. Three weak 
night-sky features are also seen extending along the entire slit. The H$\alpha$ 
and [N~II] lines show little velocity structure beyond turbulent motions in the 
ionized gas. 

Immediately to the left of the H$\alpha$ line is the He~II 6560 \AA\ line. 
This feature may have been identified by 
\citet{gm84}
as a possible signature of high-velocity gas prior to the discovery of 
the nebular He~II emission. The He~II 
emission extends only about 14 arcsec (20 pixels) along the slit. Closer 
inspection of the 6560 \AA\ line reveals that the feature is curved: at 
positions close to the star the He~II line is redshifted by about 10 km 
s$^{-1}$ with respect to the end positions of the feature. This velocity 
structure is not observed in either H$\alpha$ nor [N~II]. 

We can compare our measured H$\alpha$ and He~II velocities with the 
CO measurements of \citet{chin97}. From our 
echellogram of N44C, we measure the wavelength of the geocoronal 
H$\alpha$ was 6562.58 \AA; the observed H$\alpha$ wavelength of 
N44C was 6569.17 A; and the observed He~II wavelength of N44C was 
6566.54 A, averaged over the 20 pixels containing the bright He~II 
emission. Correcting for earth and solar motions, we find that
V(helio) = 300 km s$^{-1}$ and V(LSR) = 284 km s$^{-1}$ for 
H$\alpha$ in N44C, V(LSR) = 289 km s$^{-1}$ for the He~II line, with
uncertainties of $\pm$2 km s$^{-1}$ in the velocities.
\citet{chin97} measured the CO emission at two positions in N44;
the measurement at 05$^h$ 22$^m$ 24.8$^s$, $-$68$^{\circ}$ 01$'$ 12$''$
is closest to Star 2, 11$''$ NE of the star. At this position, the
CO velocity is V(LSR) = 281.4 km s$^{-1}$. Thus, the the H$\alpha$ 
velocity of N44C, 284 km/s, is very similar to the CO velocity, 
while the He~II emission appears to be redshifted with respect to
the molecular cloud.
We speculate that the He~II emission represents material that is being 
accelerated by the wind of the O star; the redshift of the gas would 
then indicate that the He~II emission arises primarily from gas behind 
the O star. We also infer from this that the densest gas in the nebula, 
associated with the face of the molecular cloud, is behind the star, 
similar to the Orion Nebula. 

\section{X-ray Imaging}

A deep, 108 ks exposure of the N44 complex was obtained in November 1997
with the High Resolution Imager (HRI) on board the ROSAT satellite. 
The HRI is sensitive to the energy range 0.1-2.0 keV; it has a 
38$'\times$38$'$ field of view and an on-axis angular resolution of 
$\approx5''$. 
These observations were obtained to study the diffuse X-ray emission from the 
N44 supershell, but can also be used to constrain the luminosity of any 
point source in N44C. Figure 7 shows a portion of the HRI image centered
on N44C; positions 1 and 2 on the image mark the G star and the O star
respectively. The image has been smoothed with a Gaussian having $\sigma$
= 2$\arcsec$. 

Figure 7 shows that there are no significant X-ray point sources in N44C.
A weak emission peak is seen approximately 3$\arcsec$ SE of the G star. 
This peak could be associated with the G star; the aspect solution of ROSAT 
is accurate to about 5$\arcsec$, and Galactic G stars can be X-ray sources. 
Unfortunately, no bright point sources exist in the field for 
astrometric calibration.  

We estimated a limit on the X-ray flux from a point source at the position
of the O star (Star 2) from this smoothed image. We used a 5$\arcsec$-radius 
aperture to account for the HRI PSF and the nominal positional
uncertainty, and two different regions for background. Neither the O star 
nor the G star could be considered detected based on this analysis. For
Star 2, the 3$\sigma$ limit for the count rate is 1.5x10$^{-4}$ counts 
s$^{-1}$ in the HRI bandpass. 

To convert this count rate to a limit on the X-ray luminosity L$_X$, we
must make a few assumptions and corrections. First, we assume that the
X-ray emission can be approximated by a 
\citet{rs77} 
thin plasma model with a temperature $>$ 10$^6$ K. We then use the conversion
from counts to energy given in the ROSAT Mission Description for a given
temperature, assuming that the PSPC count rate is three times that of the 
HRI. Assuming a distance of 50 kpc to the LMC and a count rate of 
4.5x10$^{-4}$ counts s$^{-1}$ in the PSPC, our computed 3$\sigma$ upper 
limits on L$_X$ are given in Table 7 as a function of temperature and
X-ray absorption N(H).

We must now estimate the opacity to X-ray emission along the line of
sight. This is typically expressed as a column density of H atoms along
the line of sight (N(H)). Note that in X-ray absorption models N(H) is 
the X-ray absorption expressed in the units per H atom using opacities
from \citet{mm83}. This absorption column density is not equal to 
the H~I column density because the absorption column density includes the 
ionized and molecular H as well. However, the H~I column density includes 
all H~I along the line of sight, while the absorption column density includes 
only the material between the object and us, so the H~I column density is
commonly used to approximate the absorption column density. We will use
this approximation as well. The H~I map of the N44C complex from 
\citet{kim98} 
indicates that the H~I column density toward N44C is only 
$\approx$ 10$^{21}$ cm$^{-2}$. The analysis of ASCA data for N44 by 
\citet{magchu96} 
indicates a similar value for N(H). Finally, we note that 
the low interstellar obscuration suggests a relatively small column of 
absorbing material in front of the nebula. Therefore, we are justified in 
assuming a absorption column density log N(H) = 21.0 - 21.5 cm$^{-2}$.

Based on parameters adopted above and the results in Table 7, we infer
that the 3$\sigma$ upper limit on L$_X$ for any currently active point 
source with T $>$ 10$^6$ K in N44C is $<$ 10$^{34}$ ergs s$^{-1}$ in the 
ROSAT 0.1-2.0 keV band. This limit may be somewhat conservative because 
the assumed absorption column density is relatively high. 
%The sigma is also high because the background is high.
Note that a softer source could have a significantly larger luminosity
and remain undetected. On the other hand, our limit is an order of
magnitude smaller than that inferred by 
\citet{pm89}, 
and is four orders of magnitude smaller than the luminosity of LMC X-1, 
which powers an X-ray ionized nebula 
\citep{pa86}.

\section{Discussion}

Our analysis of N44C still leaves us with a puzzle: where is the evidence
for the source of the He$^+$-ionizing photons that power the He~II zone?
\citet{gkcs91} 
outlined several possible mechanisms which could
account for the He~II emission. The data presented here rule out virtually
all of the proposed mechanisms. Our GHRS spectrogram of Star 2 shows it to
be consistent with a main sequence dwarf star, and thus it is incapable 
of ionizing the He~II zone according to state-of-the-art stellar atmosphere 
models. Except for, possibly, nitrogen, the nebular abundances are consistent
with average LMC values. Evidence for a previous evolved companion star is
thus largely absent. We see no evidence for high-velocity H$\alpha$ anywhere 
in N44C, ruling out the possibility that a shock front produces the He~II 
emission.  

The one possibility that remains open is the hypothesis by 
\citet{pm89} 
that N44C is a fossil X-ray-ionized nebula. They conjectured
that N44C once contained an X-ray binary similar to LMC X-1 in which the
X-ray emission had switched off some time within the recent past. Noting
the presence of He~II emission and the absence of [Ne~V] emission, they
computed the recombination time scales for both species and estimated 
that the putative X-ray source would have had to shut off between 20
and 100 years prior to 1986. 

These estimates by Pakull \& Motch were obtained assuming an electron
density of 200 cm$^{-3}$. However, we know from the present data that
$n_e$ $\approx$ 1000-2000 cm$^{-3}$ in the He~II-emitting region of 
the nebula. At this density the recombination time-scales for He$^{+2}$
and Ne$^{+4}$ are approximately 20 and 5 years, respectively. Given 
that our CTIO spectrogram was obtained in 1991, this implies that the
putative X-ray source shut down sometime between 1970 and 1978 (when 
the $Einstein$ observatory surveyed X-ray sources in the LMC). This 
is a very narrow window for the shut-off to occur. There is a further 
problem: the presence of considerable amounts of Ne$^{+3}$,
which requires 63 eV photons to exist. The [Ne~IV] 2420 \AA\ line
is strong and well-measured, as seen in Figure 3, so observational
uncertainties are small. If the collision strength for this transition
is greatly underestimated we would overestimate the Ne$^{+3}$ abundance,
but the abundance would still be significant if overestimated by a 
factor of ten. The radiative recombination coefficient for Ne$^{+3}$, 
9.8$\times$10$^{-12}$ cm$^3$ s$^{-1}$ at T = 10,000 K 
\citep{gould78}, 
is even larger than that for Ne$^{+4}$, 
and we derive a recombination timescale of only about 3 yrs for Ne$^{+3}$.
This analysis neglects dielectronic recombination, which has a much
larger cross section than radiative recombination for Ne$^{+3}$ and
Ne$^{+4}$ at nebular temperatures 
\citep{ns87}. 
Thus, if the lack of [Ne~V] emission is due to rapid recombination in 
a fossil X-ray-ionized nebula, we should not expect to 
see [Ne~IV] emission either. What keeps the Ne$^{+3}$ ionized? 

Nevertheless, as we noted earlier, the inferred L($\lambda$4686) in our 1991 
observations is much smaller than that inferred from the 1986 observations.
This suggests that we are observing the recombination of He$^{+2}$ 
with time. In addition, the inferred Q(H$^0$) for Star 2 from our data 
is much higher than that expected from an O7V star, suggesting that there 
may have been an additional source of ionizing photons. Further spectral 
monitoring of N44C is warranted to confirm if the nebula is evolving as 
expected under the fossil nebula hypothesis.

\citet{markclark75} reported the detection
by $OSO$-7 of a source, LMC X-5, near the position of N44C, but the 
positional error circle ($\approx$0.3 degree) was so large that a 
definitive identification can not be established. This source was
detected significantly (4$\sigma$ level) only in the 3-10 keV band.
\citet{grifsew77} claimed to have
confirmed this detection with the $Ariel$ observatory, but the
positional uncertainty is much larger. The candidate LMC X-5 was
not detected by \citet{lhg81} 
in their $Einstein$
observatory survey of the LMC; they placed an upper limit of 
2$\times$10$^{35}$ ergs s$^{-1}$ for the 0.15-4.5 keV luminosity
of the source. No source was detected in our ROSAT HRI observation in
1997, nor in a shorter pointed HRI observation in 1993. Pakull \& 
Motch noted that there are some cases of ``permanent'' X-ray binaries 
for which the X-ray emission subsides for periods of several years 
\citep{jfl73, psgg86}, 
although these cases are low-mass X-ray binaries, not massive stars such 
as the case here. In the low-mass X-ray binaries, the mechanism for X-ray
turn-off is proposed to be pulsations in the secondary star which 
cause the secondary to alternately fill and underfill its Roche lobe,
leading to periodic X-ray activity. This phenomenon is less likely to
occur in an O-type star. Another possibility is that an unseen compact
companion may be in a highly eccentric orbit about Star 2. Such a
companion could trigger activity through accretion of material from
the O star as the companion approaches periastron; X-ray activity would 
then shut off as the companion moves outward on its orbit. A configuration
such as this would have to be young enough that tidal forces have not
significantly circularized the orbits. An interesting possibility is
that the system is related to the Be X-ray binaries 
\citep{rvdh82}, 
although Star 2 itself does not currently show evidence 
of Be activity. 
X-ray and optical spectral monitoring, and a radial 
velocity study, are needed to constrain the models for this system. 

\section{Summary}

We have presented new CTIO and HST spectroscopy of the He~II-emitting
nebula LMC-N44C, a low-resolution UV spectrogram of its ionizing star,
new nebular kinematic data, and deep high-resolution X-ray imaging 
of the nebula. We summarize our analysis here:

(1) The GHRS spectrogram of the ionizing star (Star 2) yields a spectral
type of about O7 for the star. The Si IV, He~II, and C~IV features do
not show P Cygni profiles, showing definitively that the star is not
a supergiant. No other stellar source is visible within 5000 AU of
the O7 star in WFPC2 images, nor is any close companion evident in 
the spectrum of the star.

(2) The abundances of He, C, O, S, Ar, and Ne in the ionized gas are
all consistent with average abundances for LMC H~II regions. Nitrogen
may show an enhancement by about a factor of two, although the correction
for unseen ionization states is large and uncertain. There is thus little
evidence for enrichment by a previous evolved companion.

(3) A long-slit echelle spectrogram in H$\alpha$ + [N~II] shows no signs
of high-velocity gas ($>$ 100 km s$^{-1}$) in N44C. Thus, high-velocity
shocks do not account for the nebular He~II emission.

(4) No X-ray point source is seen in a 108 ks ROSAT HRI image of N44C.
We set an upper limit of 10$^{34}$ erg s$^{-1}$ for the 0.1-2.0 keV
luminosity of any point source in the nebula.

(5) Based on new measurements of the electron density in the He~II
emitting region, we derive recombination timescales of $\approx$ 20 
yrs for He$^{+2}$ and $\approx$ 4 yrs for Ne$^{+4}$, about a factor
of five smaller than the values derived by 
\citet{pm89}. 
This places severe constraints on the time window within which a
putative X-ray source could have turned off. An additional puzzle
is why [Ne~IV] emission is still observed in the nebula if the 
ionizing source has shut off.

(6) X-ray monitoring and optical spectral monitoring of the nebula, 
and a radial velocity study of the ionizing star, are needed to fully
understand the source of the hard ionizing photons that produced
the He~II and [Ne~IV] emission in N44C. 

\acknowledgments
Denise Taylor and Tony Keyes of STScI deserve many thanks for their 
assistance in setting up the {\it HST} observation sequence, and in 
scheduling the repeat observations before the removal of the FOS 
from {\it HST}. We also thank Kim Venn and Danny Lennon for helpful 
discussions of the nature of the O star and for providing
a high-dispersion spectrum of Star 2. Zoltan Levay provided assistance 
in producing Figure 5. Finally, we thank the referee, Nolan Walborn, 
for a number of helpful suggestions. Support for this program was 
provided by NASA through grant GO-6623-95A from STScI. 
DRG also acknowledges support from NASA-LTSARP grant NAG5-7734.

\clearpage

\begin{figure}
\vspace{16.0cm}
%\plotone{Fig1.ps}
\includegraphics{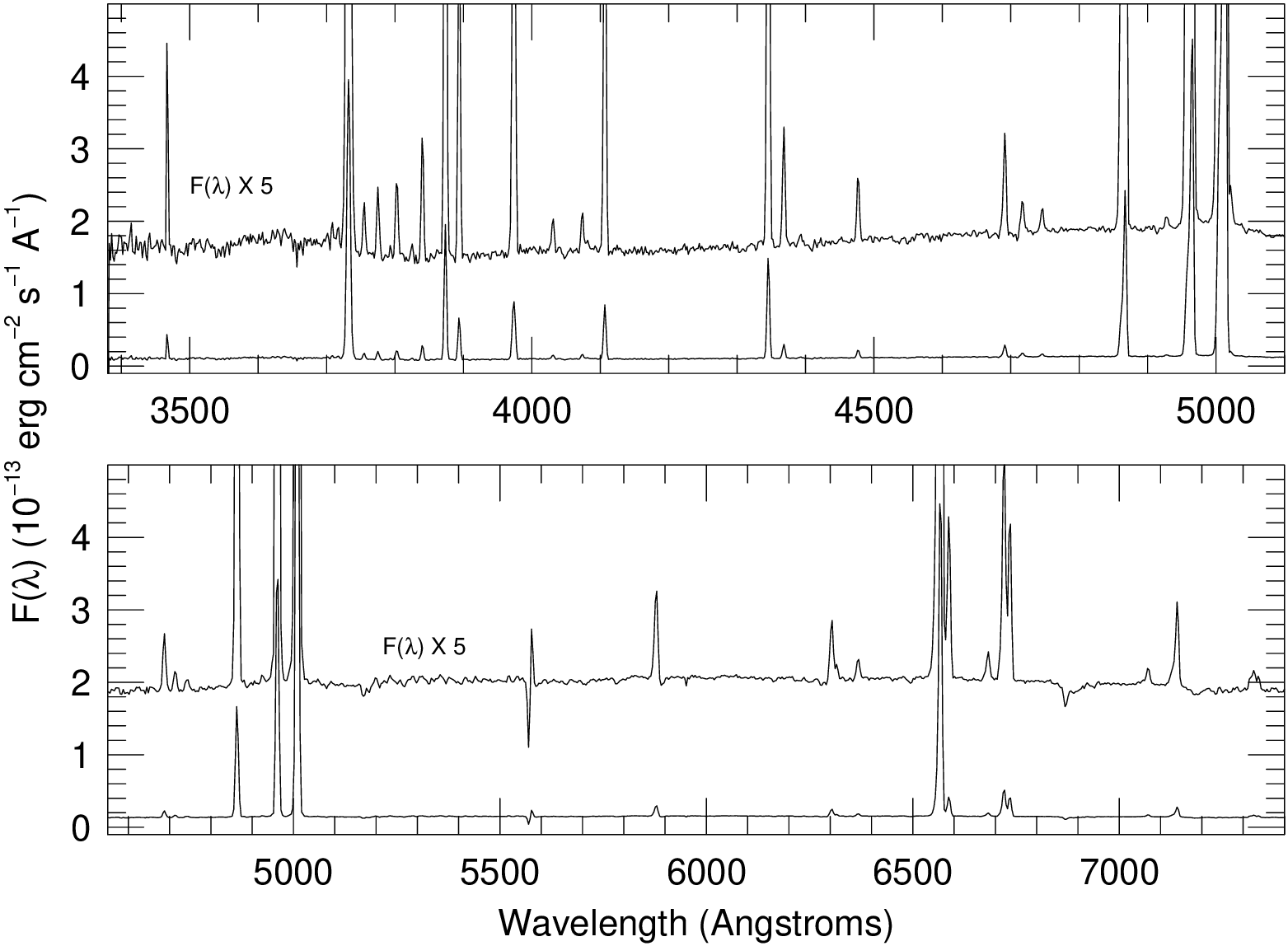}
%\plotfiddle{Fig1.ps}{6in}{000}{070}{070}{-250}{000}
\caption{Top panel: Blue CTIO spectrogram of LMC-N44C. Bottom panel: 
Red CTIO spectrogram of LMC-N44C. The upper plot in each panel is the
same spectrum expanded by a factor five in the Y scale to show
faint emission lines. }
\end{figure}

\clearpage

\begin{figure}
\vspace{16.0cm}
%\plotone{Fig2.ps}
\includegraphics{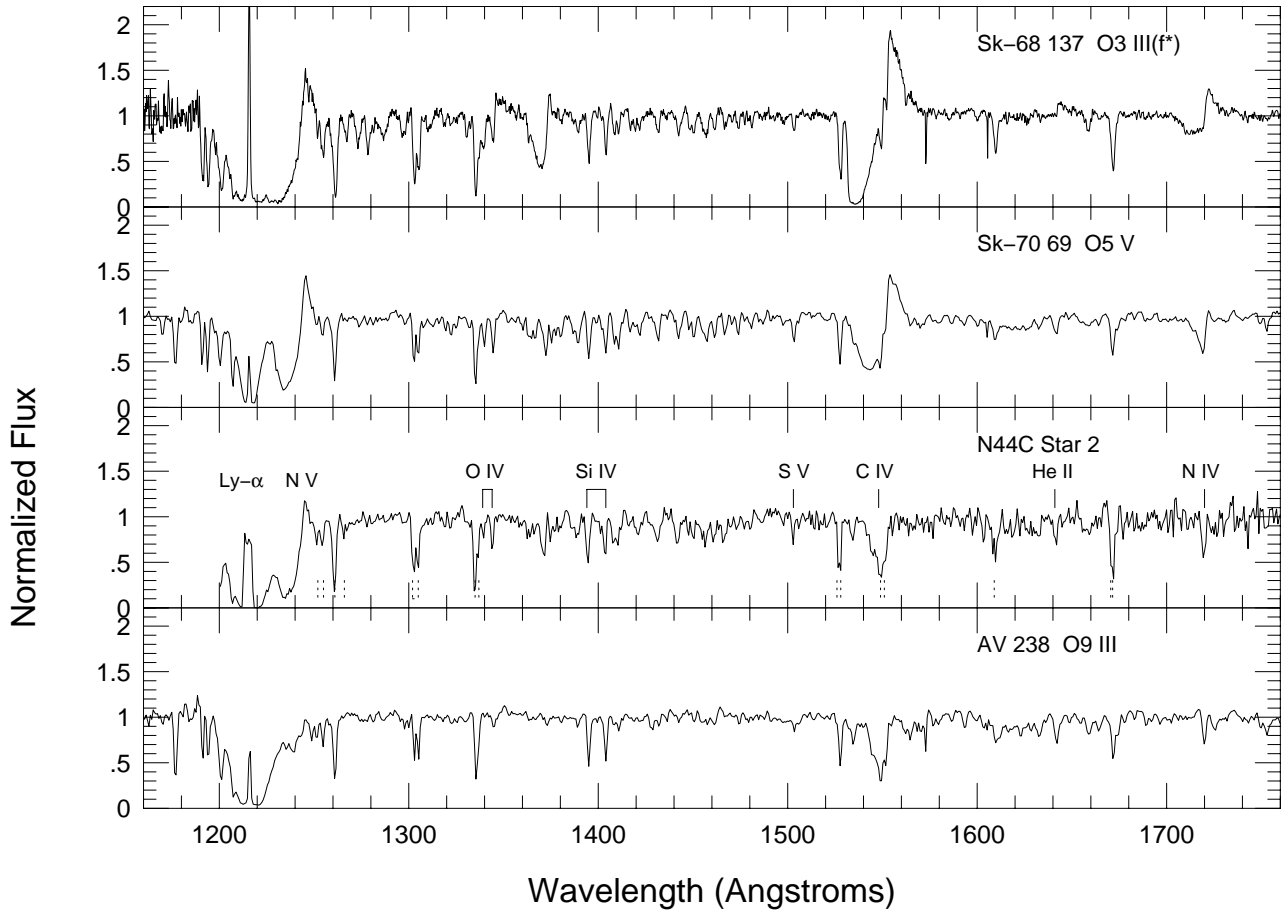}
%\plotfiddle{Fig2.ps}{6in}{000}{070}{070}{-250}{000}
\figcaption{Normalized GHRS spectrogram of Star 2, the ionizing star in 
LMC-N44C, and three comparison O III-V stars in the Magellanic Clouds. 
The Star 2 data have been smoothed with a five-point median filter. Some 
stellar features are identified above the Star 2 spectrum. The most prominent 
interstellar absorption features are marked by the dotted lines below 
the spectrum. Note that the Si IV lines may have associated interstellar 
features that cannot be distinguished at this resolution.} 
\end{figure}

\clearpage

\begin{figure}
\vspace{16.0cm}
%\plotone{Fig3.ps}
\includegraphics{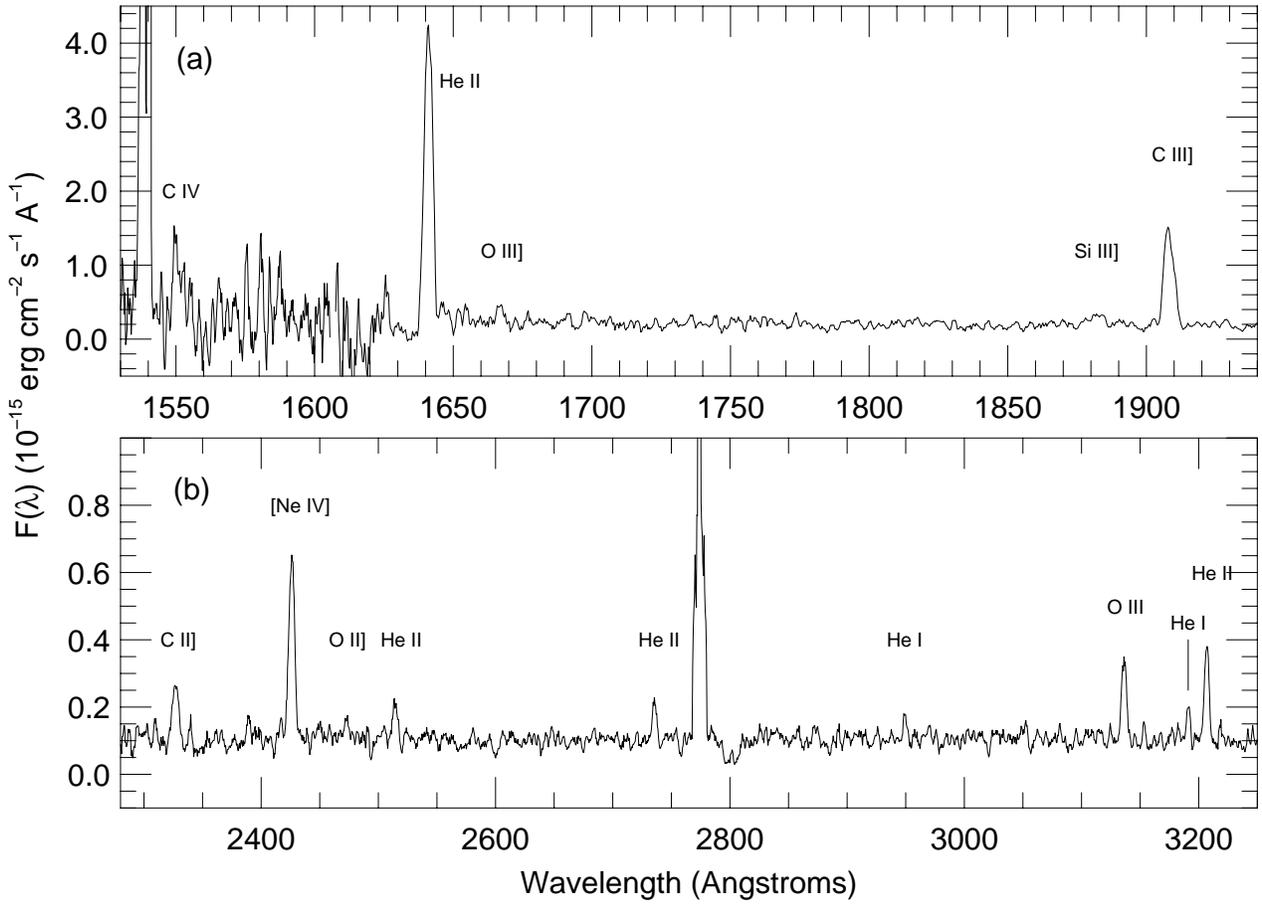}
%\plotfiddle{Fig3.ps}{6in}{000}{070}{070}{-250}{000}
\figcaption{Portions of the FOS UV spectrogram of LMC-N44C. The spectrum has 
been smoothed with a three-point boxcar. Selected emission lines are
labeled. The strong emission features near 1535 \AA\ and 2775 \AA\ 
are spurious. The vertical scales in the two panels are not the same.} 
\end{figure}

\clearpage

\begin{figure}
\vspace{16.0cm}
%\plotone{Fig4.ps}
\includegraphics{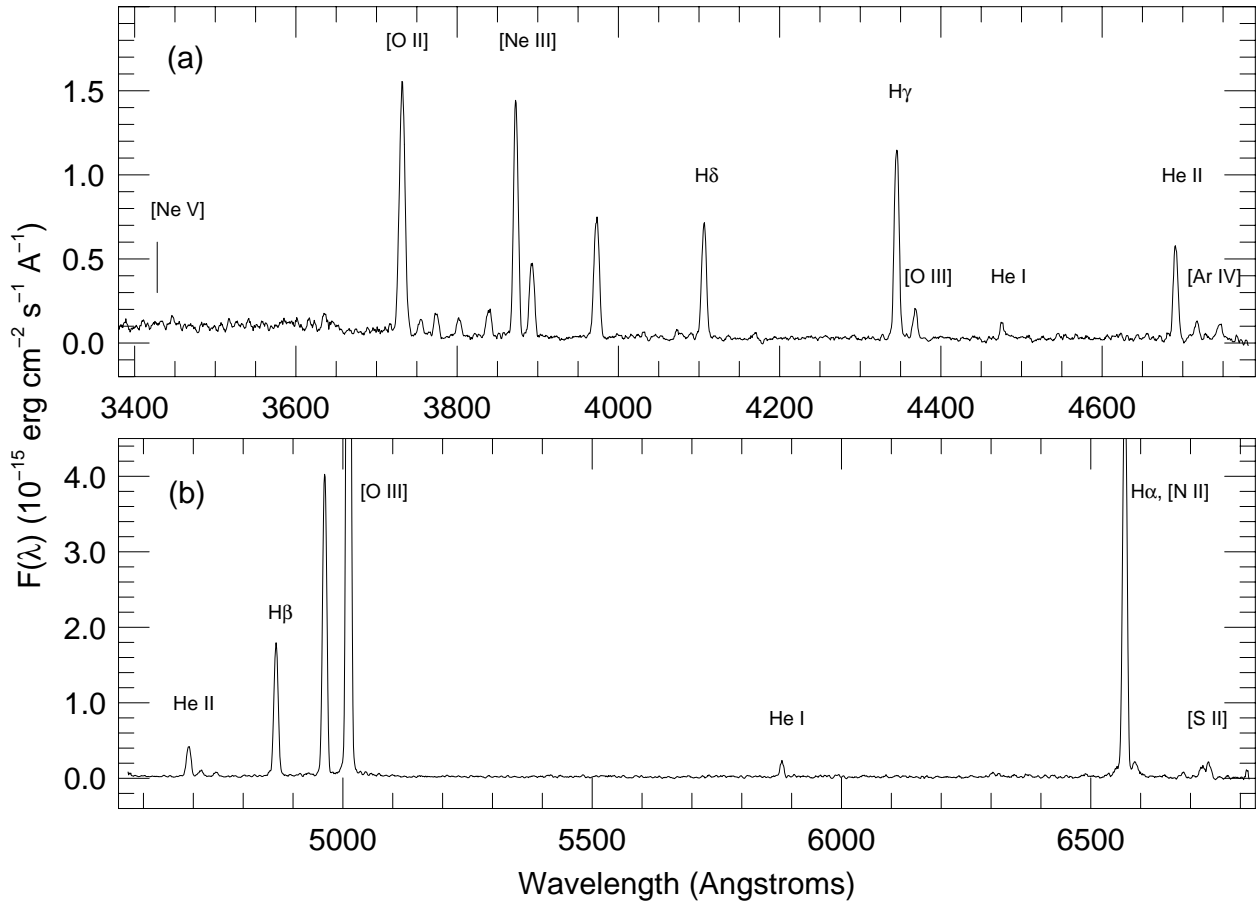}
%\plotfiddle{Fig4.ps}{6in}{000}{070}{070}{-250}{000}
\figcaption{Portions of the FOS optical spectrogram of LMC-N44C. The spectrum 
has been smoothed with a three-point boxcar. Selected emission lines are
labeled. The vertical scales in the two panels are not the same.} 
\end{figure}

\clearpage

\begin{figure}
\vspace{19.0cm}
%\plotone{Fig5.ps}
%\special{psfile="Fig5.ps" angle=00 vscale=100 hscale=100 voffset=-100 hoffset=-60}
%\plotfiddle{Fig5.ps}{6in}{000}{090}{090}{-270}{-090}
\figcaption{Color composite WFPC-2 image of LMC-N44C. North is up and 
east is to the left. Chip WF2 is at the upper left and the PC is
at the upper right. Here blue corresponds to the F502N filter ([O~III]), 
green to F656N (H$\alpha$), and red to F547M (emission-free continuum). 
A logarithmic intensity scale was used. Star 2 is located at the center
of WF2. The Wolf-Rayet star Br 25 is the bright star at bottom center 
of the PC frame, to the right of the dark dust lane.} 
\end{figure}

\clearpage

\begin{figure}
\vspace{18.0cm}
%\plotone{Fig6.ps}
\includegraphics{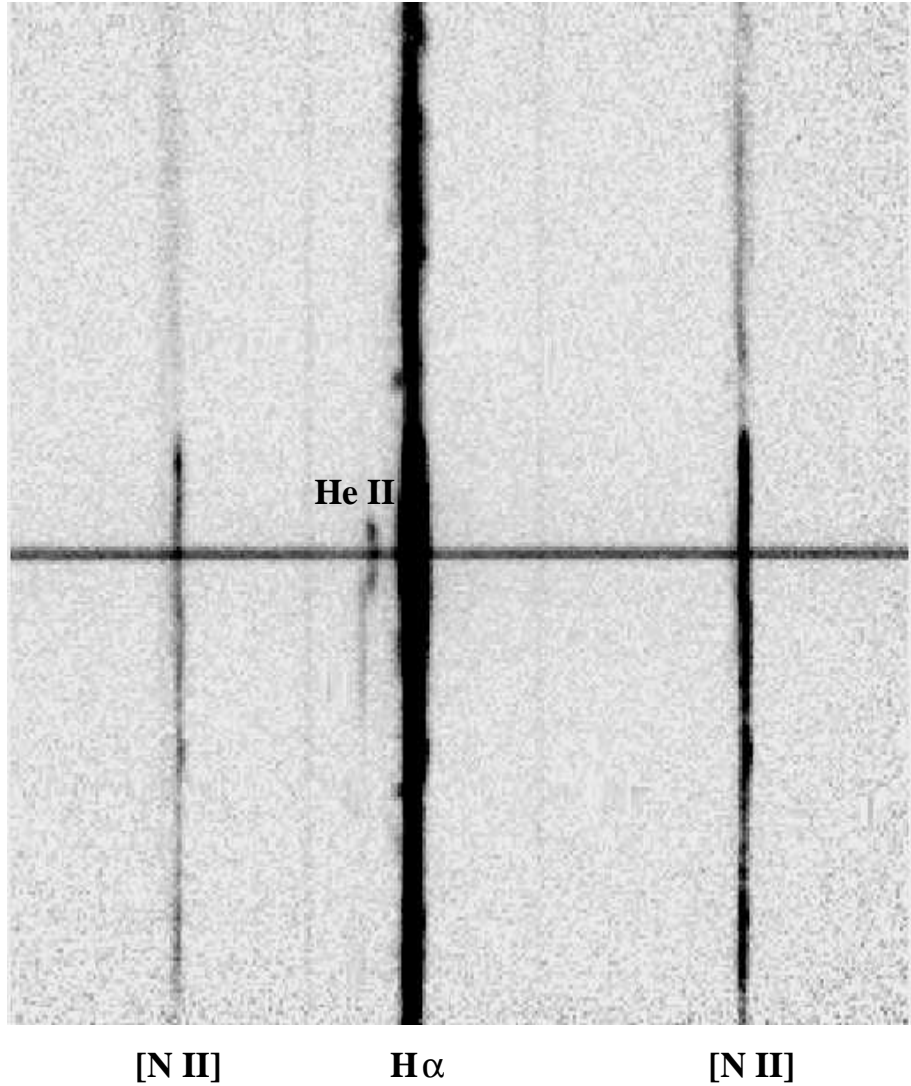}
%\plotfiddle{Fig6.ps}{6in}{000}{100}{100}{-160}{050}
\figcaption{CTIO 4-meter echelle image of the H$\alpha$ + [N II] region
of N44C. The wavelength coverage is 56 \AA\ along the horizontal axis 
and the spatial extent is 180$\arcsec$ along the vertical axis. 
East is up and bluer wavelengths are to the left.
The nebular emission lines are labeled; the 
fainter emission lines are atmospheric night-sky emission. The dark 
horizontal band is continuous emission from Star 2. } 
\end{figure}

\clearpage

\begin{figure}
\vspace{18.0cm}
%\plotone{Fig7.ps}
\includegraphics{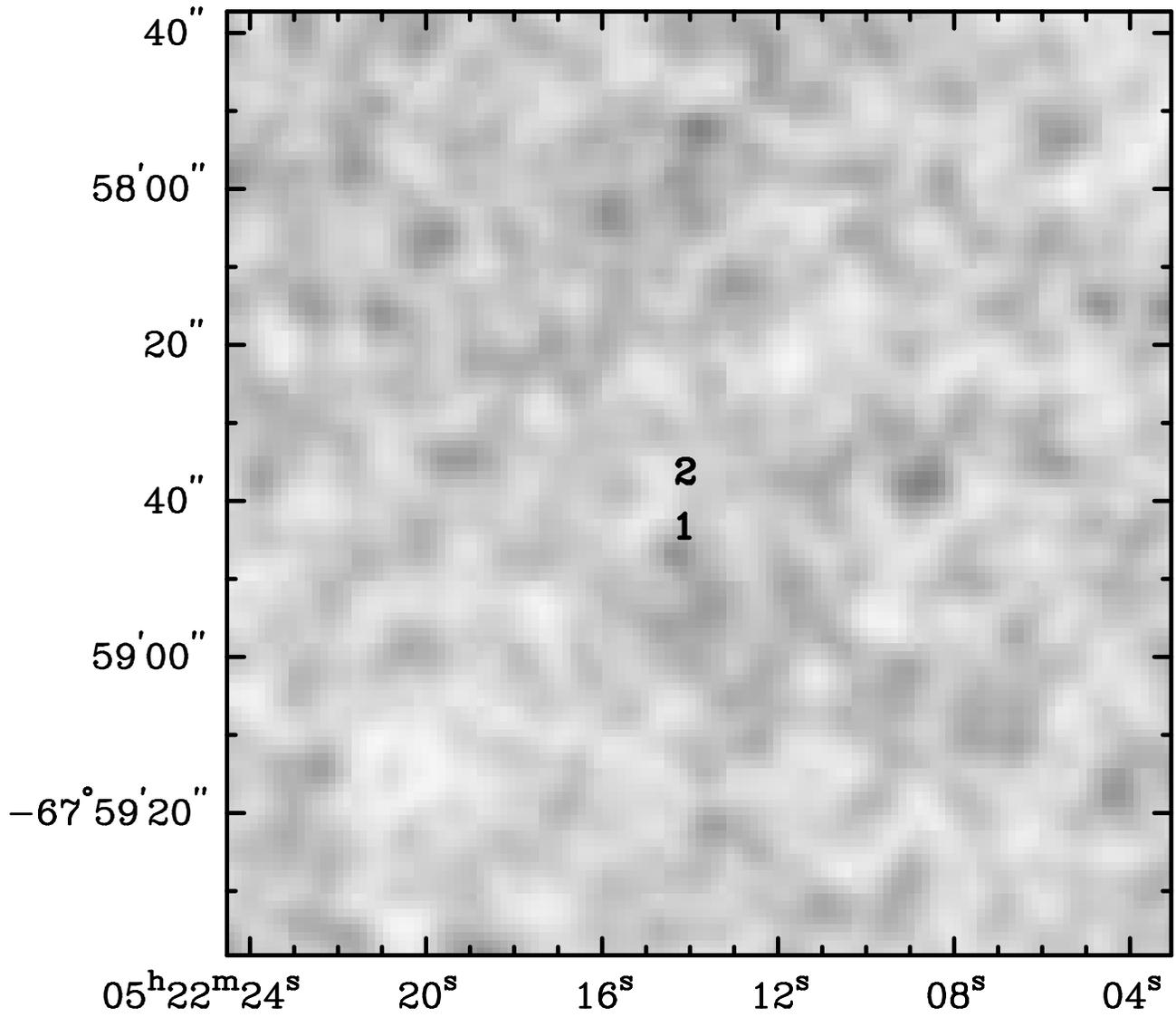}
%\plotfiddle{Fig7.ps}{6in}{000}{090}{090}{-280}{-090}
\figcaption{ROSAT HRI image of the region centered on N44C Star 2. The
image has been smoothed by a Gaussian with $\sigma$ = 2$\arcsec$.
The positions of Stars 1 and 2 are marked on the image. } 
\end{figure}

\clearpage

\begin{deluxetable}{lcccc}
\tablewidth{35pc}
\tablecaption{Journal of HST Science Observations for LMC-N44C}

\tablehead{
\colhead{Observation ID} & \colhead{Obs. Date} & \colhead{Instrument} & \colhead{Spectral Element} & \colhead{Exposure Time} }
\startdata
Z3IQ0304T & 13 Nov 1996 & GHRS     & G140L(1197-1487) &  235s\tablenotemark{a}  \\
Z3IQ0305T & 13 Nov 1996 & GHRS     & G140L(1469-1765) &  653s  \\
          &             &          &                  &        \\
Y3IQ0306T & 13 Nov 1996 & FOS/BLUE & G400H            &  360s\tablenotemark{b}  \\
Y3IQ0307T & 13 Nov 1996 & FOS/BLUE & G130H            & 1500s\tablenotemark{b}  \\
Y3IQA301T & 13 Nov 1996 & FOS/RED  & G270H            &  650s  \\
Y3IQA302T & 13 Nov 1996 & FOS/RED  & G570H            &  110s  \\
Y3IQA303T & 13 Nov 1996 & FOS/RED  & G190H            & 1200s  \\
Y3IQ5305T & 10 Feb 1997 & FOS/BLUE & G130H            & 1610s  \\
Y3IQ5306T & 10 Feb 1997 & FOS/RED  & G190H            & 1300s  \\
Y3IQ5307T & 10 Feb 1997 & FOS/RED  & G570H            &  180s  \\
Y3IQ5308T & 10 Feb 1997 & FOS/RED  & G400H            &  370s  \\
          &             &          &                  &        \\
U3IQB301T & 13 Nov 1996 & WFPC2    & F656N            &  700s  \\
U3IQB302T & 13 Nov 1996 & WFPC2    & F656N            &  700s  \\
U3IQB304T & 13 Nov 1996 & WFPC2    & F675W            &   80s  \\
U3IQB305T & 13 Nov 1996 & WFPC2    & F675W            &   80s  \\
U3IQB306T & 13 Nov 1996 & WFPC2    & F502N            &  700s  \\
U3IQB307T & 13 Nov 1996 & WFPC2    & F502N            &  700s  \\
U3IQB308T & 13 Nov 1996 & WFPC2    & F502N            &  700s  \\
U3IQB309T & 13 Nov 1996 & WFPC2    & F547M            &   80s  \\
U3IQB30AT & 13 Nov 1996 & WFPC2    & F547M            &   80s  \\
\enddata
\tablenotetext{a}{Exposure shortened due to loss of guide star lock}
\tablenotetext{b}{No signal; FOS aperture door closed during exposure}
\end{deluxetable}

\clearpage

\begin{deluxetable}{lclcccl}
\tablecolumns{7}
\scriptsize 
\tablewidth{0pc}
\tablecaption{Emission Line Intensities -- CTIO Spectroscopy} 

\tablehead{ \colhead{Ion} & \colhead{$\lambda$} & \colhead{
I($\lambda$)/H$\beta$\tablenotemark{a}} & \colhead{} & \colhead{Ion} & 
\colhead{$\lambda$} & \colhead{I($\lambda$)/H$\beta$\tablenotemark{a}} 
\\
\colhead{} & \colhead{(\AA )} & \colhead{} & \colhead{} & \colhead{} & 
\colhead{(\AA)} & \colhead{} 
}
\startdata
\ [Ne V]   & 3425 & $<$ 0.01     & & [Ar IV]  & 4740 & 0.017(0.002) \\
\ [O II]   & 3727 & 1.87(0.07)   & & [O III]  & 4959 & 2.21(0.07)   \\
\ H 12     & 3750 & 0.035(0.004) & & [O III]  & 5007 & 6.60(0.20)   \\
\ H 11     & 3770 & 0.039(0.005) & & He I     & 5876 & 0.113(0.005) \\
\ H 10     & 3798 & 0.052(0.005) & & [O I]    & 6300 & 0.067(0.003) \\
\ H 9      & 3835 & 0.069(0.007) & & [S III]  & 6312 & 0.017(0.002) \\
\ [Ne III] & 3868 & 0.61(0.02)   & & [O I]    & 6363 & 0.023(0.002) \\
\ He I     & 4026 & 0.019(0.003) & & H$\alpha$& 6563 & 2.86(0.010)  \\
\ [S II]   & 4068 & 0.021(0.003) & & [N II]   & 6583 & 0.161(0.06)  \\
\ H$\delta$& 4101 & 0.25(0.01)   & & He I     & 6678 & 0.031(0.002) \\
\ H$\gamma$& 4340 & 0.469(0.015) & & [S II]   & 6716 & 0.233(0.009) \\
\ [O III]  & 4363 & 0.063(0.003) & & [S II]   & 6731 & 0.163(0.006) \\
\ He I     & 4471 & 0.040(0.003) & & [Ar V]   & 7005 & $<$ 0.002    \\
\ He II    & 4686 & 0.064(0.003) & & He I     & 7065 & 0.017(0.001) \\
\ [Ar IV]  & 4711 & 0.019(0.002) & & [Ar III] & 7135 & 0.095(0.004) \\
\ [Ne IV]  & 4725 & $<$ 0.007    & & [O II]   & 7325 & 0.043(0.003) \\
\enddata
\tablenotetext{a} {Corrected for reddening as described in text. 1$\sigma$
uncertainties given in parentheses; upper limits are 2$\sigma$ bounds.}
\end{deluxetable} 

\clearpage

\begin{deluxetable}{lclcccl}
\tablecolumns{7}
\scriptsize 
\tablewidth{0pc}
\tablecaption{Emission Line Intensities -- FOS Spectroscopy}

\tablehead{ \colhead{Ion} & \colhead{$\lambda$} & \colhead{
I($\lambda$)/H$\beta$\tablenotemark{a}} & \colhead{} & \colhead{Ion} & 
\colhead{$\lambda$} & \colhead{I($\lambda$)/H$\beta$\tablenotemark{a}} 
\\
\colhead{} & \colhead{(\AA )} & \colhead{} & \colhead{} & \colhead{} & 
\colhead{(\AA)} & \colhead{} 
}
\startdata
\ C IV     & 1548 & $<$ 0.2       & & H I      & 3835 & 0.064 (0.005) \\
\ He II    & 1640 & 1.4 (0.1)     & & [Ne III] & 3868 & 0.55 (0.02)   \\
\ O III]   & 1667 & 0.07 (0.03)   & & H$\delta$& 4101 & 0.28 (0.01)   \\
\ N III]   & 1750 & $<$ 0.05      & & H$\gamma$& 4340 & 0.48 (0.02)   \\
\ Si III]  & 1883 & 0.06 (0.01)   & & [O III]  & 4363 & 0.063 (0.004) \\
\ C III]   & 1908 & 0.47 (0.04)   & & He I     & 4471 & 0.031 (0.004) \\
\ C II]    & 2327 & 0.05 (0.01)   & & He II    & 4686 & 0.22 (0.01)   \\
\ He II    & 2386 & 0.022 (0.009) & & [Ar IV]  & 4711 & 0.034 (0.005) \\
\ [Ne IV]  & 2424 & 0.22 (0.02)   & & [Ar IV]  & 4740 & 0.026 (0.005) \\
\ [O II]   & 2470 & 0.023 (0.008) & & [O III]  & 4959 & 2.17 (0.09)   \\
\ He II    & 2511 & 0.043 (0.007) & & [O III]  & 5007 & 6.9 (0.3)     \\
\ He II    & 2733 & 0.032 (0.005) & & He I     & 5876 & 0.088 (0.007) \\
\ He I     & 2945 & 0.013 (0.005) & & [O I]    & 6300 & 0.020 (0.006) \\
\ O III    & 3133 & 0.064 (0.006) & & [S III]  & 6312 & 0.017 (0.006) \\
\ He I     & 3188 & 0.023 (0.006) & & H$\alpha$& 6563 & 2.84 (0.12)   \\
\ He II    & 3203 & 0.078 (0.006) & & [N II]   & 6583 & 0.124 (0.008) \\
\ [O II]   & 3727 & 0.73 (0.03)   & & He I     & 6678 & 0.021 (0.007) \\
\ H I      & 3750 & 0.029 (0.005) & & [S II]   & 6716 & 0.088 (0.008) \\
\ H I      & 3770 & 0.055 (0.007) & & [S II]   & 6731 & 0.112 (0.008) \\
\ H I      & 3798 & 0.044 (0.006) & &          &      &               \\ 
\enddata
\tablenotetext{a} {Corrected for reddening corresponding to A(V) = 0.26$\pm$0.05
as described in text. 1$\sigma$ uncertainties are given in parentheses; upper
limits are 2$\sigma$ bounds.}
\end{deluxetable} 

\clearpage

\begin{deluxetable}{lllcl}
\tablecolumns{5}
\scriptsize 
\tablewidth{0pc}
\tablecaption{Derived Electron Temperatures and Densities}

\tablehead{\colhead{Spectrum} & \colhead{Diagnostic Ratio}
& \colhead{Observed Line Ratio} & \colhead{Quantity} & \colhead{Derived Value}
}

\startdata
CTIO & [S II]  6717/6731 & \phn\phn\phn1.43$\pm$0.08  & $n_e$ & $<$ 160 cm$^{-3}$      \\
CTIO & [Ar IV] 4711/4740 & \phn\phn\phn 1.12$\pm$0.18 & $n_e$ & \phn2800$\pm$2200 cm$^{-3}$ \\
CTIO & [O III] 5007/4363 & \phn143$\pm$7         & $T_e$ & 11200$\pm$400 K         \\
CTIO & [O II]  3727/7325 & \phn\phn 44$\pm$3        & $T_e$ & 10700$\pm$1200 K        \\
%CTIO & [S II]  4070/6725 & \phn\phn\phn 5.3$\pm$2.1   & $T_e$ & $\approx$ 8800 K       \\
FOS  & [S II]  6717/6731 & \phn\phn\phn 0.79$\pm$0.09 & $n_e$ & \phn1500$\pm$800 cm$^{-3}$  \\
FOS  & [Ar IV] 4711/4740 & \phn\phn\phn 1.19$\pm$0.29 & $n_e$ & \phn2000$\pm$2000 cm$^{-3}$ \\
FOS  & [O III] 5007/4363 & \phn144$\pm$10        & $T_e$ & 11200$\pm$500  K        \\
FOS  &  O III] 1666/5007 & \phn\phn\phn 0.010$\pm$0.004 & $T_e$ & 10800$\pm$800  K     \\
FOS  & [O II]  2470/3727 & \phn\phn\phn 0.03$\pm$0.01   & $T_e$ &  \phn9000$\pm$1600 K \\
\enddata
\end{deluxetable} 

\clearpage

\begin{deluxetable}{lcc}
\tablecolumns{3}
\scriptsize 
\tablewidth{0pc}
\tablecaption{Nebular Ionic Abundances} 

\tablehead{
\colhead{Ion} 
& \multicolumn{2}{c}{Abundance relative to H$^{+}$} 
\\
\colhead{} 
& \colhead{CTIO} 
& \colhead{FOS}   
}

\startdata
He$^{+}$   &  (8.4$\pm$0.3)e-2   &  (5.9$\pm$0.4)e-2    \\ 
He$^{+2}$  &  (5.3$\pm$0.3)e-3   &  (1.8$\pm$0.2)e-2    \\ 
C$^{+}$    &  \nodata            &  (4.8$\pm$1.0)e-6    \\ 
C$^{+2}$   &  \nodata            &  (5.4$\pm$0.6)e-5    \\ 
C$^{+3}$   &  \nodata            &  $<$ 1.8e-5          \\ 
N$^{+}$    &  (2.4$\pm$0.1)e-6   &  (1.9$\pm$0.1)e-6    \\ 
N$^{+2}$   &  \nodata            &  $<$ 2.7e-5          \\ 
O$^{+}$    &  (4.5$\pm$0.2)e-5   &  (2.2$\pm$0.1)e-5    \\ 
O$^{+2}$   &  (1.61$\pm$0.05)e-4 &  (1.67$\pm$0.08)e-4  \\ 
Ne$^{+2}$  &  (4.3$\pm$0.2)e-5   &  (3.9$\pm$0.2)e-5    \\ 
Ne$^{+3}$  &  $<$ 2.3e-4         &  (2.2$\pm$0.2)e-5    \\ 
Ne$^{+4}$  &  $<$ 6.9e-7         & \nodata              \\   
S$^{+}$    &  (6.6$\pm$0.4)e-7   &  (4.5$\pm$0.4)e-7    \\ 
S$^{+2}$   &  (2.6$\pm$0.3)e-6   &  (2.6$\pm$0.9)e-6    \\ 
Ar$^{+2}$  &  (6.8$\pm$0.3)e-7   &  \nodata             \\ 
Ar$^{+3}$  &  (2.6$\pm$0.3)e-7   &  (3.8$\pm$0.7)e-7    \\ 
Ar$^{+4}$  &  $<$ 3.0e-8         &  \nodata             \\ 
\enddata
%\tablenotetext{} {Upper limits based on upper limit line fluxes.}
\end{deluxetable} 

\clearpage

\begin{deluxetable}{llcc}
\tablecolumns{4}
\scriptsize 
\tablewidth{0pc}
\tablecaption{Total Element Abundances in LMC-N44C} 

\tablehead{
\colhead{Element} & \colhead{Log Abundance} &\colhead{Dufour 1990}
& \colhead{Garnett 2000}
}

\startdata
He/H  &  --1.08$\pm$0.03  & --1.07  & \nodata \\ 
C/H   &  --4.16$\pm$0.06  & --4.07  & --4.1   \\ 
N/H   &  --4.9$\pm$0.1    & --5.03  & --5.1   \\ 
O/H   &  --3.68$\pm$0.02  & --3.57  & --3.6   \\ 
Ne/H  &  --4.22$\pm$0.05  & --4.36  & --4.4   \\ 
S/H   &  --5.4$\pm$0.1    & --5.15  & --5.3   \\ 
Ar/H  &  --2.3$\pm$0.1    & --5.8   & --5.8   \\ 
\enddata
\end{deluxetable} 
\clearpage

\begin{deluxetable}{llll}
\tablewidth{0pc}
\tablecaption{X-ray Luminosity Limits for N44C-Star 2}
\tablehead{
\colhead{log N(H)} & \colhead{T = 10$^6$ K} &\colhead{T = 5$\times$10$^6$ K} 
& \colhead{T =  10$^7$ K}
}
\startdata
20.5  & $<$ 6.8$\times$10$^{33}$ ergs s$^{-1}$ & $<$ 1.7$\times$10$^{33}$ ergs s$^{-1}$  & $<$ 1.5$\times$10$^{33}$ ergs s$^{-1}$ \\
21.0  & $<$ 4.5$\times$10$^{34}$ & $<$ 3.3$\times$10$^{33}$ & $<$ 1.9$\times$10$^{33}$ \\
21.5  &    large      & $<$ 9.0$\times$10$^{33}$ & $<$ 3.6$\times$10$^{33}$ \\
22.0  &    large      & $<$ 6.7$\times$10$^{34}$ & $<$ 2.3$\times$10$^{34}$ \\
\enddata
\end{deluxetable}

\end{document}